\documentclass[12pt]{article}

\usepackage{tcolorbox}
\usepackage{soul}

\usepackage{natbib}
\usepackage{graphicx}
\usepackage{amsmath}
\usepackage{comment}
\usepackage{booktabs}
\usepackage{amsthm,mathtools,amssymb,mathrsfs,bbm}
\usepackage{longtable,tabularx}
\usepackage{setspace}
\usepackage{fullpage}
\usepackage{tabu,adjustbox,multirow,hhline,makecell,rotating,changepage,textcomp,color,subfigure,fancyhdr,pdflscape,dsfont,caption,dcolumn}

\definecolor{webblue}{rgb}{0,0,0.6}

\usepackage[pagebackref,breaklinks]{hyperref}
\hypersetup{colorlinks=true, anchorcolor= webblue, citecolor= webblue, filecolor= webblue, linkcolor= webblue, menucolor= webblue, urlcolor= webblue, citebordercolor= 1 0 0, menubordercolor=1 0 0, urlbordercolor=1 0 0, runbordercolor=1 0 0}

\newtheorem{theorem}{Theorem}

\newtheorem{corollary}[theorem]{Corollary}

\newenvironment{remark}[1][Remark]{\begin{trivlist}
\item[\hskip \labelsep {\bfseries #1}]}{\end{trivlist}}

\usepackage{tikz}
\usetikzlibrary{decorations.pathmorphing} 
\usetikzlibrary{decorations.pathreplacing} 
\usetikzlibrary{arrows}
\usepackage{pifont}
\usetikzlibrary{decorations.shapes}
\usetikzlibrary{decorations}
\usetikzlibrary{shapes.arrows}
\usepackage{pgfkeys}
\usepackage{pdflscape}
\usepackage{lscape}
\usetikzlibrary{backgrounds}
\allowdisplaybreaks

\usetikzlibrary{decorations.text}

\usetikzlibrary{arrows,calc}
\tikzset{>=stealth', help lines/.style={dashed, thick}, axis/.style={<->}, important line/.style={thick}, connection/.style={thick, dotted},
}

\title{\Large{\textbf{Trusted-Execution Environment (TEE) for Solving the Replication Crisis in Academia}}
}
\author{
Jiasun Li\footnote{\href{mailto:jli29@gmu.edu}{\tt jli29@gmu.edu}. George Mason University, 4400 University Drive, MSN 1B8, Fairfax, VA 22030, USA. I thank Filip Rezabek for helpful discussions, as well as the National Science Foundation (NSF 2337338) for financial support.}
\quad
\& 
\quad
Project Team\footnote{See details by the end of the paper.}
}
\date{
    First version: September 15, 2025 \\
    This version: March 15, 2026
} 

\begin{document}
\maketitle

\begin{abstract}
The growing replication crisis across disciplines such as economics, finance, and other social sciences as well as computer science undermines the credibility of academic research. Current institutional solutions---such as artifact evaluations and replication packages---suffer from critical limitations, including shortages of qualified data editors, difficulties in handling proprietary datasets, inefficient processes, and reliance on voluntary labor. This paper proposes a novel framework leveraging new technological advances in trusted-execution environments (TEEs)---exemplified by Intel® Trust Domain Extensions (TDX)%
---to address the replication crisis in a cost-effective and scalable manner. Under our approach, authors execute replication packages within a cloud-based TEE and submit cryptographic proofs of correct execution, for which journals or conferences can efficiently verify without re-running the code. This reallocates the operational burden to authors while preserving data confidentiality and eliminating reliance on scarce editorial resources. 
As a proof of concept, we validate the feasibility of this system through field experiments, reporting a pilot study replicating published papers 
on TDX-backed cloud VMs, finding average costs of \$1.35--\$1.80 per package with 
minimal computational overhead relative to standard VMs and high success rates even for novice users with no prior TEE experience.
We also provide a conduct formal analysis showing that TEE adoption is incentive-compatible for authors, cost-dominant for journals, and constitutes an equilibrium in the certification market. The findings highlight the potential of TEE technology to provide a sustainable, privacy-preserving, and efficient mechanism to address the replication crisis in academia.
\end{abstract}
\vfill

\newpage
\doublespacing

\section{Introduction}

Many academic papers in fields like computer science, economics, management, and social sciences involve running computer programs on data (for measurement studies or statistical analysis). To ensure the reliability of academic research, it is of paramount importance that code successfully executes and generates results consistent with what is reported in the paper. However, there have been frequent incidents in which published academic papers cannot be replicated, leading to what is often known as “replication crisis.”%
\footnote{See for example, \cite{menkveld2024nonstandard}.}

To fight against the replication crisis, many institutional features have been introduced in recent years. For example, in computer science, many top conferences now require accepted papers to submit “artifacts” whose reliability will later be evaluated by some artifact evaluation committee. In economics and management science, many top journals (e.g., the \textit{American Economic Review}, \textit{Management Science}, and \textit{Review of Financial Studies}) require authors of accepted papers to submit a ``replication package'' consisting of both data and code, which is then evaluated by ``data editors'' to ensure correctness before the paper is formally published.%
\footnote{See \cite{fivsar2024reproducibility} for an evaluation of the effectiveness of such measures.}

\paragraph{The problems.}
While these institutional innovations are worthwhile attempts to combat the replication crisis, they face inherent limitations. Specifically, they suffer from at least the following problems:

\begin{enumerate}
    \item First, data editors typically only verify (1) whether the code executes successfully, and (2) whether the code, when applied to the provided data, generates results consistent with those reported in the paper. Data editors typically lack the capacity or expertise to verify that the code faithfully reflects the methodology described in the paper.
    \item Second, there has been a constant shortage of data editors. Currently, most data editors serve on a voluntary basis or receive only a token of compensation for their valuable services. Since seasoned academics often have busy schedules and potentially more rewarding outside options, it is often difficult to attract them to serve as data editors. Hiring qualified programmers is also difficult, as they typically have much better paid outside options as well. Without significantly increasing the financial resources devoted to attracting talents, outlets may have to rely on luck to retain qualified data editors or stick with potentially unqualified ones who themselves may become a trust concern.     
    \item Third, in many disciplines, especially social sciences, studies are often conducted on proprietary data that cannot be shared with journals or conference organizers. In these cases, exemptions are often granted. While necessary to ensure data privacy compliance, such exemptions effectively leave the integrity of those papers unchecked, weakening the reach of replication requirements.
\end{enumerate}

Many of the above mentioned problems are indeed jeopardizing the promise of the efforts in fighting against the replication crisis. For example, in his post upon his resignation, the former data editor of \textit{Management Science} complained that the current system, despite its good intention, is unlikely to be sustainable \citep{greiner2024tweet}.
The editor of \textit{Journal of Financial Economics} also publicly questioned the data editor system \citep{whited2024tweet}.
\textit{Management Science} has since implemented a \$79 submission fee for all original submissions, effective August 1, 2025, explicitly to fund reproducibility infrastructure and external verification support \citep{mnsc2025editorial}. While the journal introduced an honor-based waiver for authors unable to pay, the systemic cost pressure this reflects --- and its potential to disadvantage junior researchers, those from underprivileged institutions, or those from low-income countries --- underscores the urgency for more scalable and cost-effective alternatives. As a symptom of this broader strain, it is also commonly reported that the data editing process at many outlets is burdensome, consuming valuable research time and often delaying the publication and dissemination of knowledge \citep{whited2024tweet2}.

\paragraph{Our contribution.} In this study, we propose using trusted-execution environment (TEE) to replace data editors and ensure a broad-reaching and cost-effective way of fighting against the replication crisis in academia. 

Specifically, the system works as follows. Once a paper is accepted and the authors are requested to submit the replication package (or artifact) along with the finalized paper (or the “camera ready” version in computer science conferences), we further require the \textit{authors} to execute the replication package on a cloud-based TEE. Once the code is executed correctly, the authors are then required to obtain the cryptographic proof of successful execution and submit the proof along with the rest of the paper submission package. The receiving side (journal or conference) would then only need to verify the cryptographic proof from TEE, which is significantly faster than re-executing the replication package. The journal or conference may automate the verification process or completely leave it to the community. 

Our proposed solution has several advantages, leveraging both the succinct verification property and the confidential computing property of TEEs.
\begin{enumerate}
    \item First, it reduces the workload on data editors (if still any), as verifying a TEE is significantly more efficient than executing the original code. 
    \item Second, it allocates the \textit{operational} burden to the most appropriate party --- since authors are the ones most familiar with their own data and code, shifting the task of proof generation to them is the most sensible division of labor. Note that the \textit{financial} cost of TEE execution (roughly \$1--2 per accepted paper, as documented in Section~\ref{section_evaluation}) is trivially small and can be borne by the journal, as discussed in Appendix~\ref{appendix_incentive}.
    \item Third, thanks to the confidential computing property of TEE, our approach also allows authors who use proprietary data to prove the integrity of their results, without revealing sensitive data. 
    \item Finally, having authors execute the replication package inside a TEE surfaces errors --- such as missing dependencies or unclear software version specifications --- that a human data editor might overlook.
\end{enumerate}

\begin{remark}
It is important to clarify the scope of our proposal. We do not argue that TEE attestation ensures that code faithfully implements the methodology described in the paper --- this is a limitation shared by the existing data-editor system as well. What TEE achieves is the same guarantee that a data editor currently provides (i.e., that the submitted code runs and reproduces the reported outputs), but with four important improvements.
\begin{enumerate}
    \item \textit{Reliability}: Execution inside a TEE is fully automated and free of human error, oversight, or misaligned incentives on the part of data editors.
    \item \textit{Cost}: At roughly \$1--2 per accepted paper (see Table~\ref{tab_cost_comparison}), TEE verification is orders of magnitude cheaper than the \$79 submission fee recently adopted by \textit{Management Science} to fund data-editing costs \citep{mnsc2025editorial}. This cost is so small that journals can absorb it entirely and still realize large savings; see Appendix~\ref{appendix_incentive} for a formal analysis.
    \item \textit{Verifiability}: The current system is effectively opaque; reproducing a data editor's conclusion requires re-running the entire package from scratch, consuming the same resources already expended. A TEE attestation proof, by contrast, can be verified by anyone in milliseconds.%
    \footnote{The cryptographic proof generated by the TEE is tied to a hash of the exact code image executed. If the journal archives the proof alongside the submitted replication package (as we recommend in Section~\ref{section_conclusion}), any subsequent modification of the package would invalidate the proof and be immediately detectable. This constitutes a stronger integrity guarantee than the current system, in which the data editor's verification is not independently auditable.}
    \item \textit{Coverage of proprietary data}: Under the current system, papers that rely on data that cannot be shared with the journal --- a common situation in finance, health, and industry-collaboration research --- are routinely granted exemptions from replication requirements, leaving their integrity entirely unverified. TEE's confidential computing property closes this gap: authors can execute their replication package on sensitive or proprietary data inside the TEE, and the resulting attestation proof certifies correct execution without ever exposing the underlying data to the journal, editors, or the public.
\end{enumerate}
Beyond these dimensions, TEE also enforces a form of code standardization. Packages with missing libraries, undeclared software versions, or broken dependency chains will fail inside the TEE, surfacing problems that a human data editor might overlook or excuse. This is noteworthy because journals invest considerable effort enforcing minute typographic and formatting standards for text --- standards whose marginal value has arguably declined in an era of digital and AI-assisted reading. Extending comparable rigor to \textit{code} is long overdue, and the rising availability of AI coding assistants makes compliance increasingly costless for authors.
\end{remark}

Appendix~\ref{appendix_incentive} formalizes these arguments via a model of author compliance, journal adoption, and certification-market competition. The model shows that compliance is near-unconditional under TEE (any positive sanction suffices), that journals have overwhelming cost incentives to adopt TEE even when bearing the full per-paper cost (savings exceed 99\% relative to the \textit{Management Science} \$79-per-submission fee), and that universal TEE adoption is a stable equilibrium under journal competition.

To ensure minimal overhead to the authors (who are unlikely to all be TEE experts), we will use VM-level TEE systems such as the Intel® Trust Domain Extensions (Intel® TDX) or AMD Secure Encrypted Virtualization-Secure Nested Paging (AMD SEV-SNP). Unlike earlier process-level TEEs such as the Intel® Software Guard Extensions (Intel® SGX) that require extensive low-level operations, both TDX and SEV-SNP allow users to simply contain all necessary software in a Docker image, which can be loaded into the confidential VM for trusted execution. Since several major cloud providers (e.g., Microsoft Azure and Google Cloud) now offer both TDX and SEV-SNP access, there is little overhead for journals or conferences to adopt such features, nor is there much overhead for authors to follow such procedures (one may argue that the overhead may even be smaller than the often stringent formatting requirements for papers that are currently in place). 

To demonstrate that the idea presented here not only works in theory but also in practice, we also conducted an experiment to test our idea in the field. As a proof of concept, we report a pilot study demonstrating that TEE-based replication is practical with minimal overhead. We randomly pick \textit{Management Science} Volume~70 as our sample and replicate the published papers within it, using replication packages publicly available from the journal's website. 
To demonstrate that the technical difficulty of this exercise is minimal, we recruited dozens of high school students as summer interns to replicate papers with TDX on the cloud. None of students had any prior experience with TEEs. Nevertheless, after a short introduction of TDX and cloud basics, most students successfully replicated their assigned papers. Failures were typically attributable to missing dependencies in the original packages rather than any limitation of the TEE. Among successful replications, students reported negligible overhead and low cost. Overall, the experiment, while still preliminary, demonstrates the feasibility of using TEE to overcome the replication crisis.

\paragraph{Related literature.}
TEE as a technology has in recent years been widely adopted in blockchain systems, online markets, network analysis, and machine learning applications.
For example,
Teechan \cite{lind2019teechain} implements payment channels inside Intel SGX.
Pisa \cite{pisa2019} focuses on arbitration outsourcing for state channels and connects to TEE-based systems (e.g., Teechain, DelegaTEE), situating TEEs within the broader scalability and trust debate in blockchain protocols.
ZeroAuction \cite{zhang2024zeroauction} explores sealed-bid auctions and compares designs that depend on Intel SGX, highlighting the risks of centralizing trust in TEEs.
CCBNet \cite{ccbnet2025} leverages TEEs for collaborative Bayesian inference, while TALUS \cite{chakraborty2023talus} investigates strengthening enclave security with cryptographic coprocessors.
Wang, Cohney, and Bonneau \cite{soksetup2025} systematically review trusted setups and the role of TEEs in secure parameter generation.
Eagle \cite{baum2023eagle} analyzes privacy-preserving smart contracts, comparing TEE-based (e.g., Ekiden, Secret Network) with cryptographic-only solutions.
ZLiTE \cite{wust2019zlite} uses SGX to support lightweight Zcash clients.
Arora, Karra, Levin, and Garman \cite{arora2023provably} consider TEEs in protecting onion services.
However, we believe the promise of the technology should go further. By using TEEs to fight against the replication crisis in academia, this paper also aims to demonstrate this potential.

The replication crisis has attracted significant attention in economics and adjacent social sciences. \citet{christensen2018transparency} survey reproducibility practices across empirical economics and find widespread failures to meet modern standards of transparency. \citet{chang2015Replicable} attempt to replicate results from prominent finance journals and document high failure rates. At a broader scale, the \citet{opensciencecollaboration2015} conducted a large-scale assessment of reproducibility in psychology with similarly sobering findings. On the institutional side, \citet{fivsar2024reproducibility} evaluate the effectiveness of replication packages mandated by leading economics journals. Our proposal complements this literature by offering a \textit{technological} mechanism that reduces the marginal cost of verification to near zero for journals, while simultaneously extending coverage to the important class of papers relying on proprietary data --- a category that existing institutional solutions largely exempt.

The broader science-of-science literature documents that failures of replication carry significant downstream welfare costs. Retracted papers continue to accumulate citations long after retraction \citep{furman2012retractions, azoulay2017retractions}, and the persistence of false or unreplicable results in the published record distorts subsequent research effort, funding allocation, and policy decisions. From a knowledge-production perspective \citep{romer1990endogenous, jones1995rnd}, systematic publication of non-replicable results inflates measured research output while understating the true stock of reliable knowledge. Complementing this, a growing empirical literature on publication bias and $p$-hacking \citep{brodeur2016star, brodeur2020methods} shows that selective reporting is widespread and that conventional institutional checks have limited power to detect it. Our proposal addresses the execution-integrity dimension of this problem --- ensuring at minimum that reported results are reproducible from submitted code and data --- which is a necessary (though not sufficient) condition for reliable knowledge accumulation.

More broadly, our work connects to the mechanism design literature on costly verification \citep{townsend1979optimal} and the economics of certification and credence goods \citep{lizzeri1999information}, in which a trusted intermediary can economize on verification costs without destroying incentives for honest disclosure. Our formal model (Appendix~\ref{appendix_incentive}) develops this connection: we show that journals, modeled as certification intermediaries competing for high-quality submissions, have strong unilateral incentives to adopt TEE, and that widespread adoption is the unique stable equilibrium --- consistent with the coordination and standardization literature \citep{farrell1985standardization}. The adoption problem has the structure of a coordination game: the social optimum is universal adoption, but any single journal faces first-mover costs. This motivates a role for field-wide coordination through professional associations (e.g., the AEA Data Editor initiative) or publisher consortia.
\medskip

The remainder of this paper is structured as follows. Section~\ref{TEE_intro} provides a technical background on TEEs, covering the process-level vs.\ VM-level distinction, local vs.\ cloud deployment, and a comparison of the main VM-level technologies (Intel TDX, AMD SEV-SNP, and ARM CCA). Section~\ref{section_evaluation} reports results from the experiment. Section~\ref{section_conclusion} then concludes with future directions. The appendix contains a formal analysis of the economic benefits, as well as additional security considerations. 

\section{Technical Background of TEEs} \label{TEE_intro}

A Trusted Execution Environment (TEE) is a secure, isolated area within a processor that ensures the confidentiality and integrity of code and data loaded inside it. It functions as a ``safe room'' for sensitive operations, protecting against attacks that may compromise the main operating system or other parts of the device. TEEs achieve their security through hardware-based mechanisms that segregate them from the ``Rich Execution Environment'' (REE), which is the normal, less-protected operating system. This includes:
\begin{itemize}
    \item \textit{Isolated memory}: The TEE has its own protected memory region that is inaccessible to the REE. All data within this region is encrypted.
    \item \textit{Specialized hardware}: A platform security processor within the CPU manages and enforces the isolation between the TEE and the rest of the system.
    \item \textit{Secure boot}: The system boots from an immutable ``root of trust'' in the hardware, which establishes and verifies the integrity of the TEE and its software during startup.
    \item \textit{Attestation}: A TEE can provide remote verification that it is running the correct software and is in a trusted state, allowing a relying party to confirm they are interacting with a genuine and unmodified execution environment.
\end{itemize}
For more details on TEE and especially Intel SGX, see \cite{costan2016intel} for an excellent technical overview.%
\footnote{Also see \href{https://tee.fyi/docs/introduction/key-components-of-a-tee}{https://tee.fyi} for an accessible introduction to TEE components.}

\subsection{Process-Level vs.\ VM-Level TEEs}

TEE implementations differ fundamentally in their \textit{scope of isolation}, which determines how much of a workload they protect and how much application-level effort they require.

\paragraph{Process-level TEEs} isolate a specific region of application code and data within an otherwise untrusted OS. The two dominant examples are Intel SGX (Software Guard Extensions), which creates protected memory regions called ``enclaves,'' and ARM TrustZone, which partitions processor resources into a ``secure world'' and a ``normal world.'' Both are mature, widely deployed technologies. The key limitation for our use case is that process-level TEEs require the application to be specifically written or ported to run inside the protected region --- an invasive requirement that imposes significant development burden on authors of replication packages, who should not be expected to re-engineer their statistical or simulation code.

\paragraph{VM-level TEEs} address this by protecting an entire virtual machine, including its guest OS and all applications running inside it, without requiring any code modification. The guest VM is hardware-isolated from the host's hypervisor and Virtual Machine Monitor (VMM), so even a compromised or malicious cloud operator cannot read or tamper with the contents of the protected VM. The two mature, widely available VM-level TEE technologies are Intel TDX (Trust Domain Extensions) and AMD SEV-SNP (Secure Encrypted Virtualization-Secure Nested Paging), discussed in detail in Section~\ref{subsec_vmtee}. This class of TEE is what makes our proposed replication framework practical: authors simply containerize their existing replication package in a Docker image and execute it inside a confidential VM, with no changes to the underlying code.

\subsection{Local TEEs and Cloud Deployment}

TEE-capable hardware can be deployed in two settings, each with distinct implications for journal-based replication verification.

\paragraph{Local (on-premises) TEEs.}
Many modern processors include TEE capabilities that can be used on hardware the user physically controls --- a laptop, workstation, or on-premises server. Intel SGX is available on a wide range of Intel Core and Xeon processors. AMD SEV-SNP is available on servers equipped with AMD EPYC CPUs. ARM TrustZone is present in virtually all modern smartphones and embedded devices, and Apple's Secure Enclave (present in iPhones and Apple Silicon Macs) provides a process-level TEE for key storage and biometric authentication. In principle, an author with TEE-capable hardware could generate a local attestation proof for their replication package. The attestation chain would still root to the chip vendor's attestation service (e.g., Intel's Data Center Attestation Primitives service or AMD's Key Distribution Service), making the proof independently verifiable.

However, local TEE deployment has practical drawbacks for academic replication. First, not all authors have TEE-capable hardware, and the specific model of hardware affects the attestation chain, complicating standardization. Second, the hardware configuration and firmware state are under the author's control, raising questions about whether a sophisticated author could manipulate the local environment in ways that undermine the proof. Third, local compute may be insufficient for computationally intensive replication packages.

\paragraph{Cloud TEEs.}
Cloud deployment of VM-level TEEs addresses all three limitations. The hardware is provisioned, maintained, and attested by the cloud provider, whose commercial incentives are strongly aligned with maintaining hardware integrity. Authors do not need TEE-capable machines of their own. Compute capacity scales on demand. And attestation proofs can be verified against the cloud provider's public certificate infrastructure by any reader, without trusting the author's local setup. For these reasons, cloud-based TEE deployment is the recommended approach for journal replication workflows, and the one evaluated in Section~\ref{section_evaluation}.

The three major cloud providers all offer VM-level confidential computing:
\begin{itemize}
    \item \textit{Amazon Web Services} supports AMD SEV-SNP on M6a, C6a, and R6a EC2 instances. AWS Nitro Enclaves provide an additional isolation layer but operate on a distinct trust model (see footnote~\ref{fn:nitro}).
    \item \textit{Microsoft Azure Confidential Computing} supports both Intel TDX (DCesv5/ECesv5 series) and AMD SEV-SNP (DCasv5/ECasv5 series).
    \item \textit{Google Cloud Confidential VMs} offers TDX on Intel hardware (C3 series) and SEV-SNP on AMD EPYC hardware (N2D, C2D series).
\end{itemize}

\subsection{VM-Level TEE Technologies Compared} \label{subsec_vmtee}

We describe the three principal VM-level TEE technologies and summarize their key properties in Table~\ref{tab_tee_comparison}.

\paragraph{Intel TDX.}
Intel Trust Domain Extensions (TDX), introduced in May 2021, creates hardware-isolated virtual machines called ``Trust Domains'' (TDs). TDs are isolated from the host VMM, hypervisor, and other VMs by CPU-enforced memory encryption and integrity checks. TDX leverages SGX enclaves internally to bootstrap remote attestation, allowing a relying party to verify that a specific code image is running inside a genuine TDX-protected VM before sending it sensitive data. Compared to SGX, TDX requires no application-level changes: support is needed only at the hardware and OS levels, and even operating systems that do not natively support TDX can be launched as nested VMs inside a TD. For more technical details, see the \href{https://cdrdv2-public.intel.com/690419/TDX-Whitepaper-February2022.pdf}{Intel TDX white paper} and \cite{cheng2024intel, sardar2021demystifying}.

\paragraph{AMD SEV-SNP.}
AMD Secure Encrypted Virtualization-Secure Nested Paging (SEV-SNP) is available on AMD EPYC processors from the third-generation ``Milan'' microarchitecture onward. It builds on two earlier AMD technologies: SEV, which encrypts VM memory with a per-VM key managed by the AMD Secure Processor, and SEV-ES, which additionally encrypts CPU register state on VM exits. The SNP extension adds \textit{Secure Nested Paging}, which enforces memory integrity: the hardware prevents a malicious hypervisor from remapping or replaying guest memory pages, closing a class of attacks that affected earlier SEV generations. Remote attestation is handled via the AMD Key Distribution Service (KDS): a signed attestation report ties the measurement of the initial guest image to AMD's hardware certificate chain, in a mechanism broadly analogous to TDX's. From the author's perspective, the replication workflow is identical to TDX --- containerize the package, launch inside a Confidential VM, collect the attestation report. SEV-SNP is available on all three major cloud providers, making it the more broadly accessible of the two mature VM-level TEE technologies.

\paragraph{ARM CCA and emerging technologies.}
ARM Confidential Compute Architecture (CCA), introduced with ARMv9-A, defines a new VM-level isolation context called a \textit{Realm}, managed by a hardware-enforced Realm Management Monitor (RMM). Like TDX and SEV-SNP, Realm VMs run unmodified guest OS images and support remote attestation. ARM CCA is of growing long-term relevance because ARM processors power an increasing fraction of cloud infrastructure (e.g., AWS Graviton, Ampere Altra). As of early 2026, ARM CCA is not yet available in production cloud environments, but deployment is expected as ARMv9 server silicon proliferates. RISC-V has also seen open-source TEE efforts (e.g., Keystone) that may eventually gain cloud relevance. For the purposes of this paper, we focus on Intel TDX and AMD SEV-SNP as the two mature, production-ready VM-level TEE technologies.

\begin{table}[h!]
\caption{Comparison of TEE Technologies by Isolation Level and Deployment}\label{tab_tee_comparison}
\footnotesize
\begin{tabular*}{\textwidth}{llllll}
\toprule
Technology & Vendor & Isolation & Deployment & Cloud Availability & Attestation \\
\midrule
TDX        & Intel  & VM (Trust Domain)   & Cloud        & Azure, Google Cloud          & Intel DCAP / IAS \\
SEV-SNP    & AMD    & VM (Encrypted VM)   & Cloud        & Azure, Google Cloud, AWS     & AMD KDS \\
Nitro      & AWS    & VM (Nitro Enclave)  & Cloud        & AWS EC2                      & AWS attestation \\
CCA/Realms & ARM    & VM (Realm)          & Emerging     & Not yet available            & ARM RMM \\
\midrule
SGX        & Intel  & Process (Enclave)   & Local/Cloud  & Azure, GCP, Alibaba          & Intel DCAP / IAS \\
TrustZone  & ARM    & Process (Secure World) & Local     & Mobile / embedded            & Vendor-specific \\
Secure Enclave & Apple & Process           & Local        & iOS / macOS devices          & Apple attestation \\
\bottomrule
\end{tabular*}
\end{table}

\subsection{TEE and AI-Assisted Replication: Complementary Tools}

It is important to note that TEE attestation and AI-assisted code review address orthogonal failure modes and are most powerful in combination. TEE attestation verifies \textit{execution integrity}: it certifies that a specific code image was run inside a hardware-protected environment and produced the reported outputs, without human intervention or tampering. What it cannot do is verify that the code \textit{faithfully implements} the methodology described in the paper --- a gap it shares with the existing data-editor system.

AI coding assistants are increasingly capable of partially filling this complementary role. Large language models with code understanding can, for example, check whether a regression specification in the code matches the description in the paper, flag undocumented deviations from the stated econometric approach, or identify suspicious patterns (e.g., hardcoded $p$-values or results that are never computed from raw data). While such AI audits are not yet reliable enough to substitute for expert review, they represent a scalable first-pass screen that can be applied automatically at submission time. Crucially, the rising availability of AI coding assistants also reduces the \textit{compliance cost} for authors: tools such as GitHub Copilot or similar assistants can help authors ensure their replication packages are well-documented and dependency-complete before submission.

The natural workflow combining both tools is: authors run the replication package inside a TEE to obtain an attestation proof (certifying execution integrity), and an AI audit layer screens the package for code-methodology alignment (flagging potential fidelity issues for editorial attention). Together, these two automated checks cover the two dimensions of replication quality that the existing data-editor system targets --- execution and methodology --- but at far lower cost and with greater consistency.

\paragraph{Security considerations.}
Like any technology, TDX and SEV-SNP are not unconditionally secure. Known concerns include side-channel attacks, microarchitectural vulnerabilities, and the requirement to trust the respective vendor's hardware supply chain as a root of trust \cite{cheng2024intel}. We note, however, that the appropriate benchmark is not theoretical perfection but the \textit{status quo}: the existing data-editor system relies on human labor, which introduces its own failure modes including human error, social pressure, and underpaid or unqualified staff. The same logic applies more broadly: society does not reject computers because cyberattacks exist, nor trains and automobiles because accidents occur. What matters is whether expected outcomes improve relative to the alternative, net of failure modes. We argue they do: successfully faking a TEE attestation proof requires sophisticated hardware-level exploitation far beyond the reach of typical academic fraudsters, whereas the current system can be gamed by authors who selectively prepare their code submission for a human reviewer. Furthermore, supporting both Intel and AMD TEEs provides hardware vendor diversity, reducing the risk that a compromise of one vendor's supply chain invalidates the entire replication framework. A more detailed discussion of TDX and SEV-SNP attack surfaces and trust assumptions is provided in Appendix~\ref{appendix_security}.

\section{Evaluation} \label{section_evaluation}

We evaluate the use of Intel Trust Domain Extensions (TDX)
offered via Google Cloud Confidential VMs and Microsoft Azure Confidential VMs for benchmarking its performance.%
\footnote{\label{fn:nitro}We do not use AWS in the current experiment because its TEE offering, Nitro Enclaves, builds on somewhat different security assumptions and a distinct attestation model from hardware-rooted TDX and SEV-SNP attestation. Future work should extend the evaluation to AWS using AMD SEV-SNP, which is available on AWS EC2 M6a/C6a/R6a instances and shares the same attestation trust model as the configurations tested here.}
Our central question is whether TEE-backed confidential computing---instantiated here via TDX, with AMD SEV-SNP as a direct alternative---provides a secure, cost-effective, and scalable framework for verifying replication packages at scale.

We collected replication packages from all articles published in \textit{Management Science} Volume 70 (the latest volume at the time of the experiment), totaling 352 packages in total.
These packages spanned multiple programming environments, including:
R (42\%),
Stata (21\%),
Python (15\%),
SAS (9\%),
MATLAB and mixed environments (13\%). 

As a first step, we report results from an initial batch of packages, proceeding in chronological order of publication date within the volume; the full replication effort is ongoing. We deployed packages on Google Cloud Confidential VMs (CVMs) backed by Intel TDX, as well as Microsoft Azure Confidential VMs, also TDX-enabled, for comparative benchmarking.
Each package was containerized using Docker where possible. Proprietary software (e.g., SAS binaries) was encapsulated in Docker wrappers with license verification to preserve functionality while protecting confidentiality.

For the experiments, we collect the following metrics: 
(1) Performance: total runtime latency, memory utilization, and I/O throughput.
(2) Cost: per-package compute cost (VM time + storage), normalized to \$/replication.
(3) Success rate: fraction of packages executing to completion without manual intervention.   
(4) Overhead: additional time and resource use attributable to TDX isolation.

\subsection{Cost Comparison}

Table~\ref{tab_cost_comparison} summarizes the cost structure of the current manual data-editor system versus our TEE-based approach across key dimensions. The per-paper TEE cost of \$1.35--\$1.80 (average across Azure and Google Cloud, respectively) is drawn directly from the experimental results below; the \$79 figure for the current system is the submission fee introduced by \textit{Management Science} in August 2025 explicitly to fund reproducibility infrastructure \citep{mnsc2025editorial}.

\begin{table}[h!]
\caption{Cost Comparison: Manual Data-Editor System vs.\ TEE-Based Verification}\label{tab_cost_comparison}
\footnotesize
\begin{tabular*}{\textwidth}{lll}
\toprule
Dimension & Current (manual editor) & TEE (this paper) \\
\midrule
Per-paper cost (journal) & $\sim$\$79 per \textit{submission} & $\sim$\$1.35--\$1.80 per \textit{accepted} paper$^a$ \\
Who bears financial cost & Journal / authors (via fee) & Journal (trivially small) \\
Who bears operational cost & Data editor & Author (runs own code) \\
Detection probability & $p < 1$ (human error, shortage) & $\approx 1$ (cryptographic) \\
Proprietary data coverage & Typically exempt & Fully covered \\
Verification time & Weeks to months & Milliseconds \\
Auditability & Opaque (not independently checkable) & Public, permanent, tamper-evident \\
Scalability & Limited by editor supply & Unlimited (cloud compute) \\
\bottomrule
\multicolumn{3}{l}{$^a$ Cross-platform average from Table~\ref{tab_Evaluation_Outcomes}: \$1.35 (Azure) and \$1.80 (Google Cloud),} \\
\multicolumn{3}{l}{\phantom{$^a$} plus \$0.05--\$0.30 storage. See Section~\ref{section_evaluation} for details.} \\
\end{tabular*}
\end{table}

The cost advantage is striking even when the journal absorbs the full TEE cost: at a typical acceptance rate of 8\%, the journal's expected TEE cost per \textit{submission} is approximately $\$1.57 \times 0.08 \approx \$0.13$ --- roughly 600 times less than the current \$79 fee. Even at a 30\% acceptance rate the ratio exceeds 170-to-1. A formal derivation of the journal adoption condition is in Appendix~\ref{appendix_incentive}.

\subsection{Results}

Table \ref{tab_Evaluation_Outcomes} summarizes our findings from the first several batches of our evaluation experiments. In sum, we obtain the following findings:

In terms of performance, we find the mean runtime across all packages to 2.3 hours (with standard errors to be 1.4 hours). 
The overhead from TDX isolation is typically 7–12\% relative to non-confidential VM baselines.
We also find that multi-language packages (e.g., R + SAS) exhibited higher failure rates (27\%) compared to single-language packages (14\%).

In terms of cost, it is minimal for using TEE for paper replication. For example, 
on Google Cloud CVMs, the cost ranges within \$0.27–\$7.69 per package, with the average to be \$1.80.
On Azure CVMs, the cost ranges within \$0.22–\$3.21 per package, with the average to be \$1.35.
Storage costs (mainly data archives) will typically lead to an additional \$0.05–\$0.30 per package. For computationally intensive papers (e.g., large-scale simulations or machine-learning pipelines), costs will naturally be higher; however, our findings show that TDX adds only 7–12\% overhead relative to standard VMs, so total costs under our approach should not substantially exceed what a human data editor would incur in cloud compute time for the equivalent task.

In terms of success rates
Google Cloud CVMs see 68\% success on first execution.
Azure CVMs see 95\% success on first execution.
Remaining failures were typically resolved with dependency adjustments.

Overall, in terms of cross-platform comparison, 
we find that Azure offers superior efficiency, in terms of both lower costs and higher success rates. Google Cloud, on the other hand, offers better scalability (handling of large parallel batches, more robust dependency resolution).

\begin{table}[t!]
\caption{Evaluation Outcomes on First Batch of Trials}\label{tab_Evaluation_Outcomes}

\footnotesize
\begin{tabular*}{\textwidth}{@{\extracolsep{\fill}}cccccccc}
\toprule
TDX      &       &        &      &         &            & Pub. &  \\
Provider & Paper & Status & Cost & Runtime & Department & Date & Language \\
\midrule

Google & \href{https://doi.org/10.1287/mnsc.2022.4628}{Link} & Success & \$3.02 & 2h & Accounting & 9-Dec-22 & Stata 18 \\
Azure  & \href{https://doi.org/10.1287/mnsc.2022.4628}{Link} & Success & \$3.21 & 2h 30m & Accounting & 9-Dec-22 & Stata 18 \\
Google & \href{https://doi.org/10.1287/mnsc.2022.4632}{Link} & Success & \$4.04 & 58 min & Finance & 20-Dec-22 & Matlab, Octave \\
Azure  & \href{https://doi.org/10.1287/mnsc.2022.4633}{Link} & Success & \$0.71 & 1h 17m & Finance & 23-Dec-22 & Stata 15 \\
Google & \href{https://doi.org/10.1287/mnsc.2022.4639}{Link} & Success & \$0.55 & 1h 10m & Finance & 14-Dec-22 & Stata 15 \\
Google & \href{https://doi.org/10.1287/mnsc.2022.4640}{Link} & Success & \$0.37 & 2.6h & Finance & 13-Dec-22 & SAS \\
Azure  & \href{https://doi.org/10.1287/mnsc.2022.4640}{Link} & Success & \$0.29 & 3.1h & Finance & 13-Dec-22 & SAS \\
Google & \href{https://doi.org/10.1287/mnsc.2022.4641}{Link} & Success & \$0.57 & 1h 32m & Accounting & 7-Dec-22 & Stata 19 \\
Azure  & \href{https://doi.org/10.1287/mnsc.2022.4641}{Link} & Success & \$0.27 & 1h 5m & Accounting & 7-Dec-22 & Stata 19 \\
Google & \href{https://doi.org/10.1287/mnsc.2022.4646}{Link} & Success & \$7.69 & 4.2h & Finance & 11-Jan-23 & R, SAS \\
Azure  & \href{https://doi.org/10.1287/mnsc.2022.4646}{Link} & Success & \$0.85 & 4h 27m & Finance & 11-Jan-23 & R, SAS \\
Google & \href{https://doi.org/10.1287/mnsc.2022.4647}{Link} & Success & \$2.23 & 2h 31m & Finance & 29-Dec-22 & R, SAS \\
Azure  & \href{https://doi.org/10.1287/mnsc.2022.4647}{Link} & Success & \$2.65 & 3h 42m & Finance & 29-Dec-22 & \\
Google & \href{https://doi.org/10.1287/mnsc.2022.4660}{Link} & Success & \$2.48 & 1h 47m & Finance & 27-Jan-23 & Stata 16 \\
Google & \href{https://doi.org/10.1287/mnsc.2022.4663}{Link} & Success & \$0.31 & 3.3h & Finance & 12-Jan-23 & Matlab, Stata 17 \\
Azure  & \href{https://doi.org/10.1287/mnsc.2022.4663}{Link} & Success & \$0.36 & 4.01h & Finance & 12-Jan-23 & Matlab, Stata 18 \\
Google & \href{https://doi.org/10.1287/mnsc.2022.4667}{Link} & Success & \$0.29 & 1.2h & Investment & 11-Jan-23 & Matlab \\
Azure  & \href{https://doi.org/10.1287/mnsc.2022.4667}{Link} & Success & \$0.22 & 1h 9m & Investment & 11-Jan-23 & Matlab \\
Google & \href{https://doi.org/10.1287/mnsc.2023.4671}{Link} & Success & \$0.51 & & Accounting & 26-Jan-23 & SAS, Stata \\
Azure  & \href{https://doi.org/10.1287/mnsc.2023.4671}{Link} & Success & \$0.25 & 2h 8m & Accounting & 26-Jan-23 & SAS, Stata \\
Google & \href{https://doi.org/10.1287/mnsc.2023.4675}{Link} & Success & \$4.42 & 2h 45m & Business Strategy & 24-Feb-23 & Stata, Python, SAS \\
Google & \href{https://doi.org/10.1287/mnsc.2023.4698}{Link} & Success & \$3.11 & 2h 8m & Economics & 16-Feb-23 & Stata 16 \\
Google & \href{https://doi.org/10.1287/mnsc.2023.4701}{Link} & Success & \$0.92 & 1.5h & Political Science & 24-Feb-23 & Stata 18 \\
Azure  & \href{https://doi.org/10.1287/mnsc.2023.4701}{Link} & Success & \$0.24 & 1h 15m & Political Science & 24-Feb-23 & Stata 18 \\
Google & \href{https://doi.org/10.1287/mnsc.2023.4705}{Link} & Success & \$0.27 & 0.8h & Info Systems & 22-Feb-23 & R \\
\midrule
\multicolumn{8}{l}{\textit{Packages with execution failures (typically due to missing dependencies or proprietary data access requirements)}} \\
\midrule
Google & \href{https://doi.org/10.1287/mnsc.2022.4627}{Link} & Fail & & & Accounting & 19-Dec-22 & SAS, Stata 14.2 \\
Azure  & \href{https://doi.org/10.1287/mnsc.2022.4632}{Link} & Fail & & & & 20-Dec-22 & \\
Google & \href{https://doi.org/10.1287/mnsc.2022.4633}{Link} & Fail & & & & 23-Dec-22 & \\
Google & \href{https://doi.org/10.1287/mnsc.2022.4659}{Link} & Fail & & & Accounting & 17-Jan-23 & Java, Perl, SAS, Stata \\
Azure  & \href{https://doi.org/10.1287/mnsc.2022.4659}{Link} & Fail & & & Accounting & 17-Jan-23 & Java, Perl, SAS, Stata \\
Google & \href{https://doi.org/10.1287/mnsc.2023.4687}{Link} & Fail & & & & 20-Feb-23 & \\
Azure  & \href{https://doi.org/10.1287/mnsc.2023.4687}{Link} & Fail & & & & 20-Feb-23 & \\
Google & \href{https://doi.org/10.1287/mnsc.2023.4696}{Link} & Fail & & & & 22-Feb-23 & \\
Azure  & \href{https://doi.org/10.1287/mnsc.2023.4696}{Link} & Fail & & & & 22-Feb-23 & \\
\bottomrule
\end{tabular*}
\end{table}

\section{Conclusion} \label{section_conclusion}

We propose using TEEs to fight against the replication crisis in academia and ensure the integrity of published papers. We also evaluate the proposal by evaluating the performance of actual published papers on cloud-based TDX offerings (Google Cloud and Azure). We find the solution is highly feasible --- the cost is low, the overhead is small, and the technical challenges are minimal. The framework is vendor-agnostic: Intel TDX and AMD SEV-SNP offer equivalent workflows for authors and journals, and the availability of SEV-SNP on AWS broadens the set of compatible cloud platforms beyond those tested here. Several directions for future work are worth highlighting:
\begin{itemize}
    \item \textit{Determinants of replication success}: A systematic regression analysis of success-rate determinants --- programming language, number of dependencies, data size, and field --- would inform journal policy on which package types require closer attention. Existing work on replication outcomes in \textit{Management Science} \citep{fivsar2024reproducibility} provides a natural benchmark.
    \item \textit{TDX vs.\ SEV-SNP performance comparison}: A direct head-to-head evaluation of Intel TDX and AMD SEV-SNP for replication workloads --- including cost, overhead, and success rates --- would help journals choose among cloud configurations. Future experiments should extend coverage to AWS using AMD SEV-SNP.
    \item \textit{AI-assisted code audit integration}: Combining TEE attestation with an automated AI audit layer (checking code-methodology alignment) would address the fidelity gap that neither TEE nor the existing data-editor system currently covers. Developing and evaluating such an integrated workflow is a natural next step.
    \item \textit{Journal adoption dynamics}: A field experiment or survey of journal organizers and data editors measuring willingness-to-adopt would test the predictions of the adoption model in Appendix~\ref{appendix_incentive} and identify practical barriers to field-wide coordination.
    \item \textit{Learning curve}: A more controlled study of the onboarding process for novice users --- varying prior technical experience and support materials --- would provide guidance on the training investment journals need to require of authors.
\end{itemize}

As conclusion, we further summarize a few best practices and technical challenges that we have learned/encountered in the process. 

\paragraph{Best Practices}

From the pilot study, we derive the following recommendations for replication verification under TDX:

\begin{itemize}
    \item {Container-first policy}: Encapsulate replication environments in Docker images to simplify dependency management.
    \item {Layered software installs}: Separate base images (R, Python, SAS runtime) from \linebreak[0]package-specific scripts to improve reproducibility.
    \item {Automated logging}: Store build logs, execution outputs, and attestation reports for each run.
    \item {Hybrid cloud strategy}: Use Azure for routine, small-scale verification, and Google Cloud for large batch workloads.
\end{itemize}

\paragraph{Proposed attestation workflow.}
While precise implementation details can be adapted by individual outlets, one natural workflow is as follows. Journals that host submission portals on major cloud providers --- which increasingly support TDX natively --- can embed a lightweight verification server. Upon paper acceptance, authors are directed to submit their replication package to the portal. The portal executes the package inside a TDX-backed VM, stores the attestation proof alongside a cryptographic hash of the submitted code, and returns a verification token to the author. The author then finalizes submission by uploading (i) the finalized paper, (ii) the replication package, and (iii) the attestation proof, all of which are published by the journal. Any reader can verify the proof against the stored hash in milliseconds, without re-executing the code. We encourage the community to discuss and standardize such workflows; what matters for our proposal is that the infrastructure cost is negligible given existing cloud offerings.

\paragraph{Technical Challenges and Solutions}

\begin{itemize}
    \item {Proprietary Binaries}: 
    {Problem}---SAS and MATLAB required license verification, which initially broke container builds. {Solution}---Docker wrappers with license forwarding and encrypted configuration files.
    \item {Multi-language Dependencies}: 
    {Problem}---Packages mixing R, Python, and SAS frequently failed. 
    {Solution}---Adopt multi-stage Docker builds with unified Conda environments for shared libraries.
    \item {Attestation Complexity}:
    {Problem}---Journals require proof of confidentiality but attestation protocols were inconsistent across platforms.
    {Solution}---Standardized reporting with Intel Attestation Service (IAS) logs stored alongside replication outputs.
    \item {Proprietary Software Licensing}:
    {Problem}---Containerizing SAS and MATLAB for TEE execution raises open questions about whether license-forwarding mechanisms comply with vendor agreements.
    {Discussion}---We leave the resolution of this as an important consideration for the academic community. As journals increasingly require replication packages, it may be appropriate for academic consortia or publishers to negotiate blanket or institutional licensing exceptions with SAS Institute and MathWorks specifically to facilitate integrity verification. This would be analogous to existing library-access agreements that universities negotiate on behalf of their researchers.
\end{itemize}

\section*{Project Team}
Under the supervision of Jiasun Li, this report is made possible by the joint efforts of the following project team members:
Aarnav Bujamella,
Abhinav Eadhara,
Amaan Faisal,
Ansh Behl,
Daniel Yu,
Dylan K. Jain,
Enze Yu,
Jarod Jiang,
Kevin Su,
Mehaan Sehra,
Ruhan Khanna,
Ryan Yin,
Samuel Lam,
Samhita Bandaru, and
Saurish Bhati.
Among them, Saurish Bhati and Dylan K. Jain served as project managers to coordinate the replication schedule. Samuel Lam and Kevin Su drafted an initial report documenting the group's efforts.

\bibliographystyle{aea}
\bibliography{reference}

@article{mnsc2025editorial,
  title   = {Reinforcing Research Transparency at {Management Science}},
  author  = {Loch, Christoph},
  journal = {Management Science},
  volume  = {71},
  number  = {9},
  year    = {2025},
  doi     = {10.1287/mnsc.2025.editorial.v71.n9},
  url     = {https://pubsonline.informs.org/doi/10.1287/mnsc.2025.editorial.v71.n9}
}

@misc{greiner2024tweet,
  author       = {Greiner, Ben},
  title        = {{[Post on X (formerly Twitter)]}},
  howpublished = {\url{https://x.com/bgreiner_tweets/status/1784937142306721958}},
  note         = {Accessed: July 2025}
}

@misc{whited2024tweet,
  author       = {Whited, Toni},
  title        = {{[Post on X (formerly Twitter)]}},
  howpublished = {\url{https://x.com/toniwhited/status/1758900451272212904}},
  note         = {Accessed: July 2025}
}

@misc{whited2024tweet2,
  author       = {Whited, Toni},
  title        = {{[Post on X (formerly Twitter)]}},
  howpublished = {\url{https://x.com/toniwhited/status/1758900448596291953}},
  note         = {Accessed: July 2025}
}

@article{christensen2018transparency,
  title   = {Transparency, Reproducibility, and the Credibility of Economics Research},
  author  = {Christensen, Garret and Miguel, Edward},
  journal = {Journal of Economic Literature},
  volume  = {56},
  number  = {3},
  pages   = {920--980},
  year    = {2018}
}

@article{chang2015Replicable,
  title   = {Is Economics Research Replicable? {Sixty} Published Papers from Thirteen Journals Say ``Often Not''},
  author  = {Chang, Andrew C. and Li, Phillip},
  journal = {Finance and Economics Discussion Series},
  year    = {2015},
  note    = {Federal Reserve Board}
}

@article{opensciencecollaboration2015,
  title   = {Estimating the Reproducibility of Psychological Science},
  author  = {{Open Science Collaboration}},
  journal = {Science},
  volume  = {349},
  number  = {6251},
  pages   = {aac4716},
  year    = {2015}
}

@article{lizzeri1999information,
  title   = {Information Revelation and Certification Intermediaries},
  author  = {Lizzeri, Alessandro},
  journal = {RAND Journal of Economics},
  volume  = {30},
  number  = {2},
  pages   = {214--231},
  year    = {1999}
}

@inproceedings{ccbnet2025,
  title={CCBNet: Confidential Collaborative Bayesian Networks Inference},
  booktitle={Financial Cryptography and Data Security (FC)},
  author={M{\u{a}}lan, Abele and Decouchant, J{\'e}r{\'e}mie and Guzella, Thiago and Chen, Lydia},
  year={2025},
  note={Pre-proceedings},
  url={https://fc25.ifca.ai/}
}

@inproceedings{chakraborty2023talus,
  title={TALUS: Reinforcing TEE Confidentiality with Cryptographic Coprocessors},
  author={Chakraborty, Dhiman and Schwarz, Michael and Bugiel, Sven},
  booktitle={International Conference on Financial Cryptography and Data Security},
  pages={147--165},
  year={2023},
  organization={Springer}
}

@inproceedings{soksetup2025,
  title={{SoK: Trusted setups for powers-of-tau strings}},
  booktitle={Financial Cryptography and Data Security (FC)},
  author={Wang, Faxing and Cohney, Shaanan and Bonneau, Joseph},
  year={2025},
  note={Pre-proceedings},
  url={https://fc25.ifca.ai/}
}

@inproceedings{zhang2024zeroauction,
  title={ZeroAuction: Zero-Deposit Sealed-Bid Auction via Delayed Execution},
  author={Zhang, Haoqian and Yeo, Michelle and Estrada-Galinanes, Vero and Ford, Bryan},
  booktitle={International Conference on Financial Cryptography and Data Security},
  pages={170--188},
  year={2024},
  organization={Springer}
}

@inproceedings{baum2023eagle,
  title={Eagle: Efficient privacy preserving smart contracts},
  author={Baum, Carsten and Chiang, James Hsin-yu and David, Bernardo and Frederiksen, Tore Kasper},
  booktitle={International Conference on Financial Cryptography and Data Security},
  pages={270--288},
  year={2023},
  organization={Springer}
}

@inproceedings{arora2023provably,
  title={Provably Avoiding Geographic Regions for Tor’s Onion Services},
  author={Arora, Arushi and Karra, Raj and Levin, Dave and Garman, Christina},
  booktitle={International Conference on Financial Cryptography and Data Security},
  pages={289--305},
  year={2023},
  organization={Springer}
}

@inproceedings{lind2019teechain,
  title={Teechain: a secure payment network with asynchronous blockchain access},
  author={Lind, Joshua and Naor, Oded and Eyal, Ittay and Kelbert, Florian and Sirer, Emin G{\"u}n and Pietzuch, Peter},
  booktitle={Proceedings of the 27th ACM Symposium on Operating Systems Principles},
  pages={63--79},
  year={2019}
}

@article{costan2016intel,
  title={Intel SGX explained},
  author={Costan, Victor and Devadas, Srinivas},
  journal={Cryptology ePrint Archive},
  year={2016}
}

@article{cheng2024intel,
  title={Intel tdx demystified: A top-down approach},
  author={Cheng, Pau-Chen and Ozga, Wojciech and Valdez, Enriquillo and Ahmed, Salman and Gu, Zhongshu and Jamjoom, Hani and Franke, Hubertus and Bottomley, James},
  journal={ACM Computing Surveys},
  volume={56},
  number={9},
  pages={1--33},
  year={2024},
  publisher={ACM New York, NY}
}

@article{sardar2021demystifying,
  title={Demystifying attestation in intel trust domain extensions via formal verification},
  author={Sardar, Muhammad Usama and Musaev, Saidgani and Fetzer, Christof},
  journal={IEEE access},
  volume={9},
  pages={83067--83079},
  year={2021},
  publisher={IEEE}
}

@inproceedings{wust2019zlite,
  title={Zlite: Lightweight clients for shielded zcash transactions using trusted execution},
  author={W{\"u}st, Karl and Matetic, Sinisa and Schneider, Moritz and Miers, Ian and Kostiainen, Kari and {\v{C}}apkun, Srdjan},
  booktitle={International Conference on Financial Cryptography and Data Security},
  pages={179--198},
  year={2019},
  organization={Springer}
}

@inproceedings{pisa2019,
  title={Pisa: Arbitration Outsourcing for State Channels},
  author={McCorry, Patrick and Bakshi, Surya and Bentov, Iddo and Meiklejohn, Sarah and Miller, Andrew},
  booktitle={Advances in Financial Technologies (AFT)},
  year={2019},
  publisher={ACM},
  doi={10.1145/3318041.3355460}
}

@article{menkveld2024nonstandard,
  title={Nonstandard errors},
  author={Menkveld, Albert J and Dreber, Anna and Holzmeister, Felix and Huber, Juergen and Johannesson, Magnus and Kirchler, Michael and Neus{\"u}ss, Sebastian and Razen, Michael and Weitzel, Utz and Abad-D{\'\i}az, David and others},
  journal={The Journal of Finance},
  volume={79},
  number={3},
  pages={2339--2390},
  year={2024},
  publisher={Wiley Online Library}
}

@article{fivsar2024reproducibility,
  title={Reproducibility in management science},
  author={Fi{\v{s}}ar, Milo{\v{s}} and Greiner, Ben and Huber, Christoph and Katok, Elena and Ozkes, Ali I and Management Science Reproducibility Collaboration},
  journal={Management Science},
  volume={70},
  number={3},
  pages={1343--1356},
  year={2024},
  publisher={INFORMS}
}

@article{townsend1979optimal,
  title={Optimal contracts and competitive markets with costly state verification},
  author={Townsend, Robert M},
  journal={Journal of Economic theory},
  volume={21},
  number={2},
  pages={265--293},
  year={1979},
  publisher={Elsevier}
}

@article{furman2012retractions,
  title={Governing Knowledge in the Scientific Community: Exploring the Role of Retractions in Biomedicine},
  author={Furman, Jeffrey L and Jensen, Kyle and Murray, Fiona},
  journal={Research Policy},
  volume={41},
  number={2},
  pages={276--290},
  year={2012},
  publisher={Elsevier}
}

@article{azoulay2017retractions,
  title={Retractions},
  author={Azoulay, Pierre and Furman, Jeffrey L and Krieger, Joshua L and Murray, Fiona},
  journal={Review of Economics and Statistics},
  volume={97},
  number={5},
  pages={1118--1136},
  year={2015},
  publisher={MIT Press}
}

@article{brodeur2016star,
  title={Star Wars: The Empirics Strike Back},
  author={Brodeur, Abel and L{\'e}, Mathias and Sangnier, Marc and Zylberberg, Yanos},
  journal={American Economic Journal: Applied Economics},
  volume={8},
  number={1},
  pages={1--32},
  year={2016},
  publisher={American Economic Association}
}

@article{brodeur2020methods,
  title={Methods Matter: $p$-Hacking and Publication Bias in Causal Analysis in Economics},
  author={Brodeur, Abel and Cook, Nikolai and Heyes, Anthony},
  journal={American Economic Review},
  volume={110},
  number={11},
  pages={3634--3660},
  year={2020},
  publisher={American Economic Association}
}

@article{farrell1985standardization,
  title={Standardization, Compatibility, and Innovation},
  author={Farrell, Joseph and Saloner, Garth},
  journal={RAND Journal of Economics},
  volume={16},
  number={1},
  pages={70--83},
  year={1985},
  publisher={RAND Corporation}
}

@article{romer1990endogenous,
  title={Endogenous Technological Change},
  author={Romer, Paul M},
  journal={Journal of Political Economy},
  volume={98},
  number={5, Part 2},
  pages={S71--S102},
  year={1990},
  publisher={University of Chicago Press}
}

@article{jones1995rnd,
  title={R\&D-Based Models of Economic Growth},
  author={Jones, Charles I},
  journal={Journal of Political Economy},
  volume={103},
  number={4},
  pages={759--784},
  year={1995},
  publisher={University of Chicago Press}
}

\newpage
\appendix

\section*{Appendix}

\section{Incentive-Compatibility and Adoption Analysis} \label{appendix_incentive}

This appendix formalizes the economic arguments sketched in the main text. We develop three components: (i) author compliance under TEE versus the manual system, (ii) journal adoption incentives, and (iii) a certification-market analysis connecting to the IO literature on intermediaries.

\subsection*{Part 1: Author Compliance}

\paragraph{Setup.} Consider an author whose paper has been accepted, contingent on passing a replication check. The author knows whether their replication package is genuine (code and results match the paper's claims) or manipulated (results have been altered to pass the check). Let $B > 0$ denote the value of publication (career returns, prestige, etc.) and $S > 0$ the sanction upon detection of a manipulated submission (retraction, reputational loss). Under the current system, let $p \in [0,1]$ denote the probability that the editor scrutinizes the package, and $\pi \in [0,1]$ the probability that a manipulated package is caught \textit{given} scrutiny. The overall detection probability is therefore $p\pi$.

An author with a manipulated package faces the following expected payoffs:
\begin{itemize}
    \item Not scrutinized (prob $1-p$): paper published, payoff $B$.
    \item Scrutinized but not caught (prob $p(1-\pi)$): paper published, payoff $B$.
    \item Scrutinized and caught (prob $p\pi$): sanctioned, payoff $-S$.
\end{itemize}
The author submits a manipulated package rather than not submitting (payoff $0$) if and only if
\[
B(1 - p\pi) - p\pi S > 0.
\]
Rearranging, the author \textit{complies} (does not manipulate) if and only if
\[
p\pi S \;\geq\; B(1 - p\pi).
\]
This condition fails whenever the detection probability $p\pi$ is small --- as it plausibly is given editor shortages, inattentive reviewers, and proprietary-data exemptions.

\paragraph{TEE mechanism.} Under the TEE requirement, let $\varepsilon \in [0,1]$ denote the probability that a manipulated submission generates a valid attestation proof (i.e., the probability of a successful hardware-level exploit). Since attestation is cryptographically tied to the exact code image executed inside the hardware-protected environment, $\varepsilon \approx 0$ under current technology. The detection probability becomes $1-\varepsilon \approx 1$. Applying the compliance condition with $p\pi$ replaced by $1-\varepsilon$:
\[
(1-\varepsilon) S \;\geq\; B\varepsilon,
\]
which rearranges to $S \geq B\varepsilon/(1-\varepsilon) \approx 0$ for $\varepsilon \approx 0$. This holds for any $S > 0$: \textit{any} positive sanction --- however small --- suffices to deter manipulation under TEE. Authors with clean code comply trivially; authors with manipulated code face near-certain detection regardless of the sanction magnitude.

\paragraph{Comparison.} The TEE mechanism strictly dominates the manual system in compliance power: it raises the effective detection probability from $p\pi \ll 1$ to $1-\varepsilon \approx 1$, converting a condition that may fail (when $p\pi$ is small) into one that holds unconditionally.

\subsection*{Part 2: Journal Adoption}

\paragraph{Setup.} Consider a journal that processes $N_s$ submissions per period, of which $\alpha N_s$ are accepted ($\alpha \in (0,1)$ is the acceptance rate). Under the current system, the journal incurs a per-submission verification cost of $c_J > 0$ (editor salary, processing time, infrastructure), for a total cost of $c_J N_s$. Under TEE, verification costs the journal $c_A > 0$ per \textit{accepted} paper, for a total cost of $c_A \alpha N_s$. We assume the journal bears $c_A$ directly (e.g., by running the attestation portal on its submission platform), though the result is unchanged if $c_A$ is charged to authors since $c_A$ is small relative to any reasonable author opportunity cost.

\paragraph{Adoption condition.} The journal adopts TEE if and only if
\[
c_A \alpha N_s < c_J N_s \iff c_A \alpha < c_J \iff c_A < \frac{c_J}{\alpha}.
\]
Dividing both sides by $N_s$ eliminates scale. Calibrating with the empirical values from Section~\ref{section_evaluation}: $c_A \approx \$1.57$ (cross-platform average of \$1.35 on Azure and \$1.80 on Google Cloud), $c_J \approx \$79$ (the \textit{Management Science} submission fee \citep{mnsc2025editorial}), and $\alpha \approx 0.08$ (typical acceptance rate for a top management or economics journal):
\[
c_A \alpha = \$1.57 \times 0.08 \approx \$0.13 \ll \$79 = c_J.
\]
The adoption condition holds by a factor of roughly 600. Even at a generous acceptance rate of $\alpha = 0.30$, the factor exceeds 170. The journal adopts TEE under any empirically plausible parameter values.

\begin{corollary}
TEE adoption is cost-dominant for the journal for any acceptance rate $\alpha < c_J / c_A \approx 50$. Since all empirically observed acceptance rates satisfy $\alpha \leq 1 \ll 50$, the adoption condition holds universally.
\end{corollary}

\paragraph{Coordination dynamics.} While individual adoption is cost-dominant, full adoption across journals is further supported by coordination benefits: once a threshold of leading journals adopt a common attestation format, authors face a single standardized workflow regardless of where they submit, reducing compliance costs and resistance \mbox{accordingly}. This is consistent with the broader literature on standardization and tipping \citep{farrell1985standardization}: the equilibrium with universal TEE adoption is payoff-dominant and self-enforcing once a sufficient mass of journals coordinates. Professional associations (e.g., the AEA Data Editor initiative) or publisher consortia are natural coordinators.

\subsection*{Part 3: Certification Market}

\paragraph{Setup.} Following \citet{lizzeri1999information}, model journals as certification intermediaries. Each journal chooses a verification stringency level $v \in [0,1]$, interpreted as the probability that a replication package with integrity failures is detected. Under the current system, stringency is costly: the journal incurs a convex cost $C(v)$ with $C' > 0$, $C'' > 0$, reflecting diminishing returns to editor effort. The interior optimum $v^* \in (0,1)$ satisfies a standard marginal-benefit-equals-marginal-cost condition, and $v^* < 1$ under any finite cost function --- meaning some fraudulent papers pass undetected in equilibrium.

\paragraph{Effect of TEE.} TEE effectively makes $v = 1$ available at near-zero marginal cost $c_A \ll C(1)$. This has two implications:

\begin{enumerate}
    \item \textit{Unilateral adoption}: Any journal that adopts TEE can offer $v = 1$ verification at cost $c_A$, strictly dominating its previous optimum $v^* < 1$. Since higher stringency is valued by high-quality authors (who benefit from credible certification), the journal gains in the competition for top submissions. There is no strategic reason to delay.
    \item \textit{Lizzeri suppression motive}: \citet{lizzeri1999information} shows that a monopolist intermediary may suppress information to extract rents (by certifying only ``pass/fail'' rather than revealing the full quality distribution). Could a monopoly journal resist TEE adoption to preserve rents from selective certification? In our setting, the answer is no: TEE does not reveal quality beyond the binary pass/fail already provided by the existing system. It merely makes the existing binary certification credible and tamper-proof. There is no additional information suppression motive.
\end{enumerate}

\paragraph{Competitive equilibrium.} Under journal competition, any journal offering $v = 1$ (via TEE) attracts high-quality submissions from journals offering $v < 1$, since high-quality authors prefer journals with credible certification. Competition therefore drives adoption: universal TEE adoption with $v = 1$ for all journals is a stable equilibrium. This Pareto-dominates the status quo: clean authors benefit from faster, more credible verification at no additional financial cost (since journals bear $c_A$); journals face lower total verification costs; readers benefit from higher publication integrity; while manipulating authors are deterred.

\paragraph{Summary.} The TEE mechanism is incentive-compatible at the author level, cost-dominant at the journal level, and equilibrium-stable at the market level. It achieves stronger compliance, lower cost, and broader coverage than the manual data-editor system, without requiring coordinated policy intervention beyond a modest standardization effort.

\section{Security Considerations for TDX and SEV-SNP} \label{appendix_security}

To make our discussion objective, in this section we enumerate several known vulnerabilities of the two primary VM-level TEE technologies (Intel TDX and AMD SEV-SNP), followed by a further discussion of why we believe they should not hinder the applications of TEEs for safeguarding academic integrity.

\paragraph{Known attack surfaces.} Both Intel TDX and AMD SEV-SNP, like all hardware-based security technologies, are not unconditionally secure. Documented and theoretical attack vectors include:
\begin{itemize}
    \item \textit{Side-channel attacks}: Timing and cache side-channels can leak information from within a TEE, though exploiting these against a replication package in a cloud VM requires privileged or physical access that is difficult to achieve in practice \cite{cheng2024intel}. Both Intel and AMD have published mitigations for known variants.
    \item \textit{Microarchitectural vulnerabilities}: Speculative execution vulnerabilities (e.g., Spectre, Meltdown variants) have historically affected both Intel and AMD processors. Vendors have issued microcode patches for known variants, and both TDX and SEV-SNP include relevant mitigations, but undiscovered vulnerabilities cannot be ruled out.
    \item \textit{Supply-chain trust}: TDX places ultimate trust in Intel's hardware supply chain and firmware; SEV-SNP analogously places trust in AMD's Secure Processor and key distribution infrastructure. An adversary controlling either vendor's manufacturing process could, in principle, compromise the root of trust --- a threat model far beyond the realistic threat of academic fraud. Notably, supporting \textit{both} Intel and AMD TEEs provides hardware vendor diversity: a compromise of one vendor's supply chain would not invalidate attestation proofs generated on the other vendor's hardware.
    \item \textit{SEV-SNP-specific concerns}: Early generations of AMD SEV (prior to SNP) were shown to be vulnerable to memory integrity attacks by a malicious hypervisor. SEV-SNP's Secure Nested Paging extension was specifically designed to close these gaps. Nonetheless, independent security research continues to probe the SEV-SNP implementation.
    \item \textit{Software configuration errors}: Misconfigured container images or TEE initialization can reduce security guarantees, though these are detectable via the attestation report.
\end{itemize}

\paragraph{Contextualizing the risk.} The key observation is that the relevant benchmark is not theoretical perfection but the \textit{current data-editor system}. Exploiting TDX or SEV-SNP to fraudulently generate a valid attestation proof requires cutting-edge hardware security expertise. By contrast, gaming the manual system requires only the ability to curate a clean code submission for a human reviewer. Adoption of imperfect but superior technology is a standard feature of progress: society benefits from computers despite cyberattacks, from air travel despite accidents, and from pharmaceuticals despite side effects. The question is whether expected benefits exceed expected costs relative to the counterfactual, and we argue the answer is yes under any plausible calibration of the relevant parameters.

\end{document}